\documentclass{mn2e}
\input epsfig.sty
\newcommand{\be}{\begin{equation}}
\newcommand{\ee}{\end{equation}}

\newcommand\lsim{\mathrel{\rlap{\lower4pt\hbox{\hskip1pt$\sim$}}
    \raise1pt\hbox{$<$}}}
\newcommand\gsim{\mathrel{\rlap{\lower4pt\hbox{\hskip1pt$\sim$}}
    \raise1pt\hbox{$>$}}}
\newcommand\esim{\mathrel{\rlap{\raise2pt\hbox{\hskip0pt$\sim$}}
    \lower1pt\hbox{$-$}}}

\begin{document}
\journal{}
\title[Cosmic reionization constraints on the nature of cosmological 
perturbations]
{Cosmic reionization constraints on the nature of cosmological 
perturbations}
\author[Pedro P. Avelino and Andrew R. Liddle]
{Pedro P. Avelino$^{1}$ and Andrew R. Liddle$^{2}$\\ 
$^1$Centro de F\'{\i}sica do Porto e Departamento de F\'{\i}sica da
Faculdade de Ci\^encias da 
Universidade do Porto,\\ \hspace*{1cm} Rua do Campo Alegre 687,
4169-007, Porto, Portugal\\
$^2$Astronomy Centre, University of Sussex, Brighton BN1 9QH, United
Kingdom} 
\maketitle
\begin{abstract}
We study the reionization history of the Universe in cosmological
models with non-Gaussian density fluctuations, taking them to have a
renormalized $\chi^2$ probability distribution function parametrized
by the number of degrees of freedom, $\nu$.  We compute the ionization
history using a simple semi-analytical model, considering various
possibilities for the astrophysics of reionization. In all our models
we require that reionization is completed prior to $z=6$, as required
by the measurement of the Gunn--Peterson optical depth from the
spectra of high-redshift quasars. We confirm previous results
demonstrating that such a non-Gaussian distribution leads to a slower
reionization as compared to the Gaussian case. We further show that
the recent WMAP three-year measurement of the optical depth due to
electron scattering, $\tau=0.09 \pm 0.03$, weakly constrains the
allowed deviations from Gaussianity on the small scales relevant to
reionization if a constant spectral index is assumed. We also confirm
the need for a significant suppression of star formation in
mini-halos, which increases dramatically as we decrease $\nu$.
\end{abstract}
%%%%%%%%%%%%%%%%%%%%%%%%%%%%%%%%%%%%%%%%%%%%%%%%%%%%%%%%%%%%%%%%%%%%%%

\section{Introduction}

The reionization of the Universe is a direct result of the formation
of luminous sources which produce the photons responsible 
(see Barkana \& Loeb 2001 for a review). Given that the first stars 
appear in low-mass halos formed at high redshift, the
reionization history of the Universe is expected to be a powerful
probe of the amplitude and nature of density fluctuations on small
scales.

The detection of Gunn--Peterson troughs (Gunn \& Peterson 1965) in the
absorption spectra of distant quasars suggests a late reionization for
the Universe, at a redshift $z \sim 6$ (Becker et al.~2001; Fan et
al.~2003; White et al.~2003; Fan et al.~2005; Gnedin \& Fan 2006).
The indications of a high optical depth from the first-year WMAP
results (Kogut et al.~2003; Spergel et al.~2003), implying early
reionization, led to a flurry of papers on quite complex ionization
scenarios (e.g.~Haiman \& Holder 2003; Hui \& Haiman 2003; Ciardi,
Ferrara \& White 2003; Cen 2003; Chiu, Fan \& Ostriker 2003;
Melchiorri et al.~2005). Simplicity has now largely been restored by
the recent more precise third-year WMAP results (Page et al.~2006;
Spergel et al.~2006), which indicate a much smaller optical depth,
$\tau = 0.09 \pm 0.03$. This is consistent both with the quasar
results and with a much simpler reionization history with a fast
transition from a neutral to a ionized universe (see for example
Choudhury \& Ferrara 2006).  Recent results by Haiman \& Bryan (2006)
suggest that this can only be achieved if star formation in
high-redshift mini-halos has been significantly suppressed (see also
Wyithe \& Loeb 2006).

In the present paper we perform a complementary study, investigating
the dependence of the reionization history of the Universe on the
nature of the cosmological perturbations, in the context of
cosmologies permitted by the WMAP three-year results.  We are
motivated by the suggestions that a non-Gaussian contribution, for
example from cosmic defects, might have been able to reconcile the
high optical depth suggested by the WMAP first-year results and a low
redshift of (complete) reionization implied by the quasar data
(Avelino \& Liddle 2004; Chen et al.~2003). With the new simpler
ionization models, non-Gaussian models should be better
constrained. 

We compute the reionization history of the Universe using a simple
semi-analytical model similar to that used by Haiman \& Bryan (2006)
(based on Haiman \& Holder 2003). We consider various possibilities
for the astrophysics of reionization, subject to the constraint that
reionization is completed prior to $z=6$ as required by the
measurement of the Gunn--Peterson optical depth from the spectra of
bright quasars at high redshift.

\section{The mass fraction}

We use the Press--Schechter approximation (Press \& Schechter 1974) to
compute the mass fraction, $f(M)$, associated with collapsed objects
with mass larger than a given mass threshold $M$.  This was originally
proposed in the context of initial Gaussian density perturbations, and
later generalized to accommodate non-Gaussian initial conditions
(Lucchin \& Matarrese 1988; Chiu, Ostriker \& Strauss 1998).  In both
cases the mass fraction is assumed to be proportional to the
fraction of space in which the linear density contrast, smoothed on
the scale $M$, exceeds a given threshold $\delta_{{\rm c}}$:
\be f(M)= A_f \int_{\delta_{{\rm c}}}^\infty {\cal P}(\delta) d
\delta \,.  
\ee 
Here ${\cal P}(\delta)$ is the one-point probability distribution
function (PDF) of the linear density contrast $\delta$, and $A_f$ is a
constant which is computed by requiring that $f(0)=1$ ($A_f=2$ in the
case of Gaussian initial conditions). This generalization of the
original Press--Schechter approximation has successfully reproduced
the results obtained from $N$-body simulations with non-Gaussian
initial conditions (Robinson \& Baker 2000).

A top-hat filter will be used to perform the smoothing 
\be
W(kR)=3\left[ \frac{\sin (kR)}{(kR)^3}-\frac{\cos
(kR)}{(kR)^2}\right]\,, 
\ee
where $M=4 \pi \rho_{{\rm m}} R^3/3$ (here $\rho_{{\rm m}}$ is the
background matter density). We take $\delta_{{\rm c}}=1.7$, motivated
by the spherical collapse model and $N$-body simulations. For
spherical collapse $\delta_{{\rm c}}$ is almost independent of the
background cosmology (e.g.~Eke, Cole \& Frenk 1996).

The evolution of the dispersion of the density field is given by  
\be
\sigma_R(z)=\frac{g(\Omega_{{\rm
m}},\Omega_\Lambda)}{g(\Omega_{{\rm 
m}}^0,\Omega_\Lambda^0)} \; \frac{\sigma_R(0)}{1+z}\,, 
\ee
where the suppression factor $g(\Omega_{{\rm m}},\Omega_\Lambda)$
accounts for the dependence of the growth of density perturbations on
the cosmological parameters $\Omega_{{\rm m}}$ and $\Omega_\Lambda$
(Carroll, Press \& Turner 1992; Avelino \& de Carvalho 1999).

We shall assume that the non-Gaussian density contrast has a
chi-squared probability distribution function (PDF) with $\nu$ degrees
of freedom, the PDF having been shifted so that its mean is zero (such
a PDF becomes Gaussian when $\nu \to \infty$). Hence,
\be
f(M) = 
\frac{Q(\nu/2,\delta_{{\rm c}}/\sigma_R \times {\sqrt
{\nu/2}}+\nu/2)}{Q(\nu/2,\nu/2)} \,,   
\ee 
where $Q(a,x)$ with $a>0$ is the incomplete gamma function defined by
\be
Q(a,x) \equiv \frac{\Gamma(a,x)}{\Gamma(a)} \,,
\ee
Here
\be
\Gamma(a,x) \equiv \int_x^\infty e^{-t} t^{a-1} dt 
\ee
and $\Gamma(a)=\Gamma(a,0)$. When $\nu \to \infty$ we have 
\be
f(M) \to {\rm erfc}\left( \frac{\delta_{{\rm c}}}{{\sqrt 2} \,
\sigma_R}\right) \,.   
\ee 

\section{The power spectrum}

Historically, one of the main uncertainties in determining the
reionization history of the Universe was the lack of an accurate
determination of various cosmological parameters. However, these
uncertainties have largely been removed by a growing body of precise
cosmological data.  Throughout, we shall adopt a cosmological model
motivated by the three-year WMAP results (Spergel et al.~2006). As
they find that WMAP alone does not require a running of the spectral
index, we will take as our base cosmology their preferred model with a
power-law initial spectrum.  The parameters are a matter density
$\Omega_{{\rm m}}^0=0.24$, dark energy density
$\Omega_\Lambda^0=0.76$, baryon density $\Omega_{{\rm B}}^0=0.042$,
Hubble parameter $h=0.73$, normalization $\sigma_8=0.74$, and
perturbation spectral index $n=0.95$.

To determine the amplitude of perturbations on the short scales
relevant to reionization, we use the transfer function from Bardeen et
al.~(1986), so that the power-spectrum is given by
\be
{\cal P}(k)  \propto k^{n} \left[ \frac{\ln(1+\epsilon_0 
q)}{\epsilon_0 q} \left({ \sum_{i=0}^4(
\epsilon_i q)^i}\right)^{-1/4}\right]^2,
\ee 
where $q=k/h \Gamma$, $[k]={\rm{Mpc}}^{-1}$,
\be
\epsilon=[2.34, 3.89, 16.1, 5.46, 6.71],
\ee 
and
\be
\Gamma=\Omega_{{\rm m}}^0 h \,,
\ee
is the shape parameter. The dispersion of the density field is given by 
\be
\sigma_R^2 = \int_0^\infty k^2 {\cal P}(k)
W^2(kR)dk\,.
\ee

For Gaussian fluctuations the Press--Schechter formalism leads to an
underestimate of the halo mass function at the high-mass end (Jenkins
et al.~2001), while the presence of baryons leads to a suppression of
power on small cosmological scales. These two opposite effects are of
the same order of magnitude and consequently we shall ignore the
baryon correction to the shape parameter (Sugiyama 1995). Also, it has
been shown that small deviations from the predicted mass function
could arise in particular in the limit of rare events (Avelino \&
Viana 2000; Inoue \& Nagashima 2002). However, since we will fix the
redshift at which reionization is to be completed by adjusting
efficiency parameters, these uncertainties will not affect our
results.

\section{The reionization model}

Following Haiman \& Bryan (2006) (see also Haiman \& Holder 2003) we
will classify dark matter halos into three different categories
according to their virial temperatures
\begin{center}
\begin{tabbing}
$4 \times 10^2\,{\rm K}$   $\lsim$ T $\lsim  10^4\, {\rm K}$ \  (Type
II) \,,\\ 
$10^4\,{\rm K}$  $\lsim$ T $\lsim 9\times 10^4\,{\rm K}$ \  (Type Ia)\,,\\
\hspace{1.15cm} T $\gsim 9\times 10^4\,{\rm K}$ \  (Type Ib)\,.
\end{tabbing}
\end{center}
The total fraction of the mass in the Universe that is condensed into
Type II, Ia, and Ib halos is given by \be F_{\rm coll,II}(z) =
f(M_{\rm II})-f(M_{\rm Ia}),
\label{eq:fii}
\ee
\be
F_{\rm coll,Ia}(z) =  f(M_{\rm Ia})-f(M_{\rm Ib}),
\label{eq:fia}
\ee 
and
\be
F_{\rm coll,Ib}(z) = f(M_{\rm Ib}),
\label{eq:fib}
\ee where the halo masses $M_{\rm II}=M_{\rm II}(T_{\rm II},z)$,
$M_{\rm Ia}=M_{\rm Ia}(T_{\rm Ia},z)$, and $M_{\rm Ib}=M_{\rm
Ib}(T_{\rm Ib},z)$ for the
virial temperatures $T_{\rm II}= 4 \times 10^2$\,K, $T_{\rm Ia}=
10^4$\,K and $T_{\rm Ib}= 9\times 10^4$\,K are obtained from 
\be
\frac{M}{10^6{\rm M_\odot}} =  \left(\frac{T}{1800 K}\right)^{3/2}
\left(\frac{1+z}{21}\right)^{-3/2}\,.  
\label{eq:mt}
\ee
These halos differ in their virial temperatures, and consequently in
the cooling processes which are effective in each case. Ionizing
sources can form in type II halos only in the neutral regions of the
intergalactic medium and provided there is a sufficient abundance of
${\rm H_2}$ molecules.  On the other hand, type I halos can cool and
form ionizing sources independently of the ${\rm H_2}$ abundance. However,
while type Ia halos can form new ionizing sources only in neutral
regions, the formation of new ionizing sources in type Ib halos can
also occur in the ionized regions of the intergalactic medium.

The recombination rate per unit of time and volume can be written as 
$\alpha_{\rm
B}C_{\rm HII} \langle n_{\rm HII} \rangle^2$, where $\alpha_{\rm B}$
is the recombination coefficient of neutral hydrogen to its excited
states ($= 2.6\times10^{-13}~{\rm cm^3~s^{-1}}$ at $T=10^4$K),
$\langle n_{\rm HII} \rangle$ is the mean number density of ionized
hydrogen, and $C_{\rm HII}\equiv \langle n_{\rm HII}^2 \rangle /
\langle n_{\rm HII} \rangle^2$ is the mean clumping factor of ionized
gas. Assuming that $C_{\rm HII}$ does not vary with redshift, the
probability $P(t_{\rm i},t)$ that a ionizing photon produced at a time
$t_{\rm i}$ is still contributing to reionization at a time $t >
t_{\rm i}$ is given by
\be
P(t_{\rm i},t)=\exp\left(\frac{t_{\rm r}}{t}-\frac{t_{\rm r}}{t_{\rm i}}
\right) \,,
\ee
where $t_{\rm r}=\alpha_{\rm B} C_{\rm HII} X \rho_B(t_{\rm i}) t_{\rm
i}^2 /m_{\rm p}$ where $\rho_{\rm B}$ is the baryon density, $X$ is
the hydrogen mass fraction and $m_{\rm p}$ is proton mass.
Alternatively we may adopt the simple power-law relation of the
evolution of the clumpiness $C_{\rm HII}$ advocated by Haiman \& Bryan
(2006)
\be
C_{\rm HII}(z)-1=9\left(\frac{1+z}{7}\right)^{-\beta}\,,
\label{eq:clump}
\ee  
which in the $\beta=0$ case reduces to $C_{\rm HII}=10$.

The HII filling factor $F_{\rm HII}(z)$ is then obtained from the
collapsed gas fractions as 
\begin{eqnarray}
\nonumber
&& F_{\rm HII}(z)= 
\int_{\infty}^{z}dz^{\prime} 
\left\{
  \epsilon_{\rm I} \frac{dF_{\rm coll,Ib}}{dz} (z^{\prime}) 
+ \left[1-F_{\rm HII}(z^\prime)\right]\right.\times\\
&&\left.\left[  \epsilon_{\rm I} \frac{dF_{\rm coll,Ia}}{dz} (z^{\prime}) 
        + \epsilon_{\rm II}  \frac{dF_{\rm coll,II}}{dz} (z^{\prime})
 \right]
\right\}P(z',z),
\label{eq:fhii}
\end{eqnarray}
where $\epsilon_{\rm I}$ and $\epsilon_{\rm II}$ are respectively the
efficiencies (number of ionizing photons produced per proton in
collapsed regions) of type I halos and type II halos (it being assumed
that the efficiencies in type Ia and Ib are equal).  The second term
on the right-hand side of equation~(\ref{eq:fhii}) includes a factor
of $(1-F_{\rm HII})$, which explicitly takes into account the fact
that new ionizing sources should appear in type II and type Ia halos
only in regions which have not yet been ionized. The injection
efficiency, $\epsilon$, of ionizing photons into the intergalactic
medium can be parametrized as $\epsilon=N_\gamma f_* f_{\rm
esc}$. Here $f_*$ is the fraction of baryons that turn into stars,
$N_\gamma$ is the average number of ionizing photons produced per
stellar proton, and $f_{\rm esc}$ is the fraction of these ionizing
photons which escape into the intergalactic medium.

Type II halos are expected to host very massive stars, which are
efficient at generating ionizing photons. On the other hand, it is
expected that only a small fraction of the available gas will be
incorporated into stars. Type I halos are less efficient at producing
ionizing photons, but that is compensated by a larger $f_*$.  In this
paper we will assume that the efficiencies do not depend on $z$ and
consider the values advocated by Haiman \& Bryan (2006) for type II 
halos, that is $N_\gamma=80000$, $f_*=0.0025$ and $f_{\rm esc}=1$,
so that $\epsilon_{\rm II}=200$ if no suppression of star formation in
type two halos occurs ($\epsilon_{\rm II}=0$ otherwise). We fix the
efficiency of type I halos such that the
Universe becomes fully ionized at $z=6.5$. This means that
$\epsilon_{\rm I}$ depends on $\nu$. However, we
find that the efficiencies are lower only by a factor of approximately
$2$ in the $\nu=1$ case compared to the $\nu \to \infty$ one. In the
Gaussian case this approximately corresponds to the parameters chosen
by Haiman \& Bryan (2006), that is $N_\gamma \sim 4000$, $f_* \sim
0.15$ and $f_{\rm esc} \sim 0.2$ for type Ia and type Ib halos so that
$\epsilon_{\rm I} \sim 120$.

The optical depth is calculated using
\begin{eqnarray}
\nonumber
 \tau(z) & = & c \sigma_{\rm T} \int_t^{t_0} dt' \ n_e(t')  \nonumber
 \\ & = &
\frac{1.08 c \sigma_{\rm T} X}{m_p} \int_z^0 dz \frac{dt}{dz}\rho_B
 F_{\rm HII} \\ & = &
\tau^* \int_0^z dz'\frac{(1+z')^2}{ \left( \Omega_{\rm m}^0 (1+z')^3 +
\Omega_\Lambda^0 \right)^{1/2}} \,  F_{\rm HII}\,, 
\end{eqnarray}
where 
\be
\tau^*=\frac{3 H_0 X \Omega^0_{\rm B} \sigma_{\rm T} c}{8\pi G m_{\rm
p}} \sim 0.056 \, \Omega^0_{\rm B} h \,.
\ee
The factor of $1.08$ approximately accounts for the contribution of 
Helium reionization. Here we assume that ${\rm HII}$ and ${\rm HeII}$ 
fractions are identical and that the ${\rm HeII}$ to ${\rm HeIII}$ 
transition takes place at $z < 6$.

\section{Results}

In Figure 1 we plot the mass fraction in type II, type Ia and type Ib
halos, as a function of redshift $z$, for chi-squared distributions
with different numbers of degrees of freedom
($\nu=1,10,100,\infty$). We clearly see that the evolution of mass
fraction with redshift becomes becomes more gradual for smaller values
of $\nu$, thus making reionization a slower process in the
non-Gaussian case. This can be confirmed in Figure 2 where we plotted
the evolution of the corresponding ionized fraction of hydrogen as a
function of redshift $z$, for two possible ionization histories (see
Figure 2).  In each case we required that reionization is completed by
$z=6.5$.  The solid line represents a reionization model with no
contribution from type II halos ($\epsilon_{\rm II}=0$), while the
dotted line represents a model where type II halos are the dominant
ionization source at high redshift ($\epsilon_{\rm II}=200$). As
expected, the slower evolution of the mass fraction in type Ia, type
Ib halos and type II halos with redshift at small $\nu$ is reflected
in the evolution of ionized fraction of hydrogen, which also becomes a
slower function of redshift.

\begin{figure}
\centering
\psfig{figure=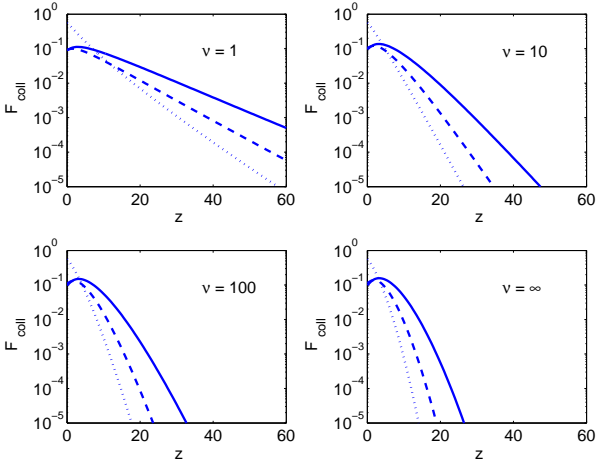,width=8cm}
\caption[Figure1]{\label{f:figure1} The evolution of the mass
fraction, $F_{\rm coll}$, in type II, type Ia and type Ib halos
(solid, dashed and dotted lines respectively), as a function of
redshift $z$, for chi-squared distributions with different numbers of
degrees of freedom ($\nu=1,10,100,\infty$).}
\end{figure}

\begin{figure}
\centering
\psfig{figure=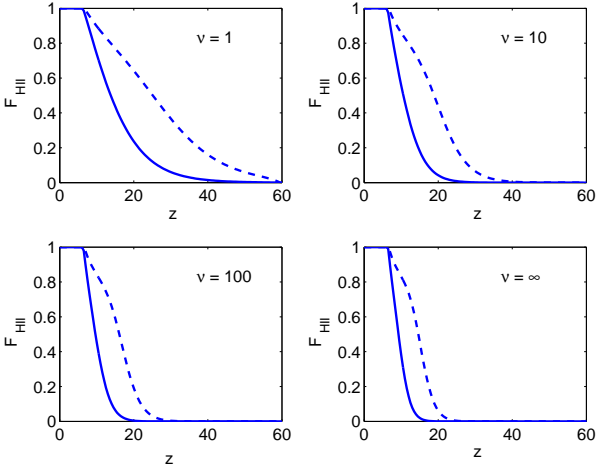,width=8cm}
\caption[Figure2]{\label{f:figure2} The evolution of the ionization
fraction of hydrogen, $F_{\rm HII}$, as a function of redshift, $z$,
for the mass fractions considered considered in Fig.~\ref{f:figure1}
and for two possible ionization histories. $\epsilon_{\rm I}$ is
chosen so that reionization is completed by $z=6.5$. The solid and
dotted lines represent reionization models with $\epsilon_{\rm II}=0$
(negligible contribution from type II halos) and $\epsilon_{\rm
II}=200$ (type II halos are the main ionization source at high
redshift) respectively. Here we assumed that the clumping factor,
$C_{\rm HII}$, is given by equation (\ref{eq:clump}) with $\beta=2$.}
\end{figure}

\begin{figure}
\centering
\psfig{figure=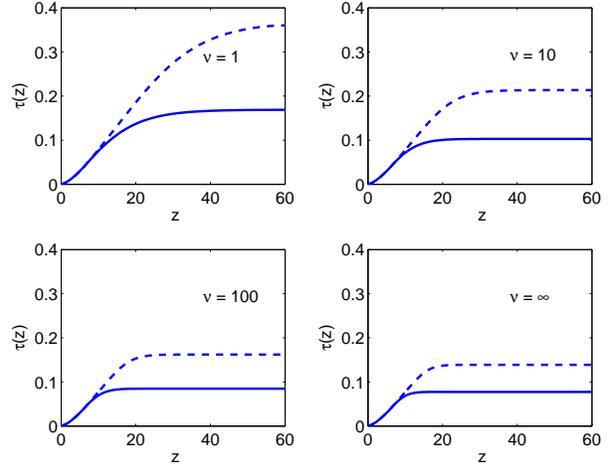,width=8cm}
\caption[Figure3]{\label{f:figure3} The optical depth due to Thomson
scattering as a function of redshift, $z$, for the models considered
in Fig.~\ref{f:figure2}.}
\end{figure}

\begin{figure}
\centering
\psfig{figure=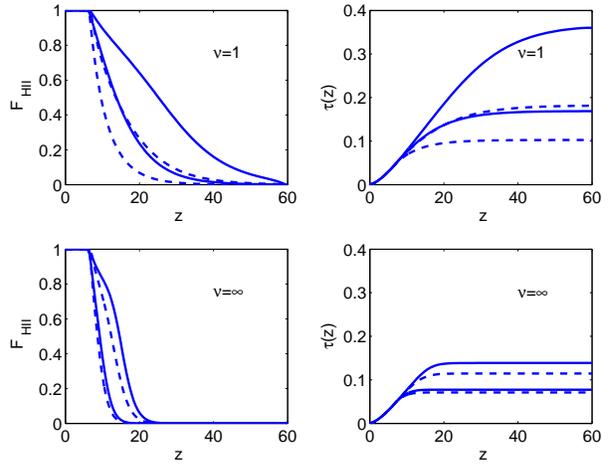,width=8cm}
\caption[Figure4]{\label{f:figure4} The evolution of the ionization
fraction of hydrogen, $F_{\rm HII}$, and the integrated optical depth
due to Thomson scattering, $\tau$, as a function of redshift, $z$, for
a model with $\nu=1$ (top) and a model with $\nu=\infty$
(bottom). Here we consider two possible ionization histories with
$\epsilon_{\rm II}=0$ (negligible contribution from type II halos) and
$\epsilon_{\rm II}=200$ (type II halos are the main ionization source
at high redshift) and two different evolutions for the clumping factor
parametrized by $\beta=2$ (solid line) and $\beta=0$ (dashed line).}
\end{figure}

In Figure 3 we plot the optical depth due to Thomson scattering as a
function of redshift, $z$. Here, we confirm the results of Haiman \&
Bryan (2006) showing that in order not to overproduce the optical
depth, star formation in mini-halos appears to have been suppressed
even if we assume Gaussian fluctuations. As we decrease $\nu$ the need
for this suppression increases dramatically. 

In Figure 4 we show the impact of two different parametrizations for
the evolution of the clumping factor ($\beta=2$ and $\beta=0$) on the
reionization history of the Universe. We see that the poor knowledge
of the redshift dependence of the clumping factor, $C_{\rm HII}(z)$,
introduces significant uncertainties in the predicted reionization
history, which become increasingly large at small $\nu$.

Finally, in Figure 5 we plot the integrated optical depth (up to
$z=60$) due to Thomson scattering, $\tau$, as a function of the number
of $\chi^2$ degrees of freedom considering different evolutions for
the clumping factor parametrized by $\beta=2$ (solid line), $\beta=1$
(dashed line) and $\beta=0$ (dotted line) and a reionization model
with a negligible contribution from type II halos ($\epsilon_{\rm
II}=0$). The horizontal stripe represents the range of $\tau$ allowed
by the WMAP three-year data (at $68 \%$ confidence level). We see that
given the model uncertainties reionization alone cannot yet rule out
any value of $\nu$ with a significant degree of confidence.

\begin{figure}
\centering
\psfig{figure=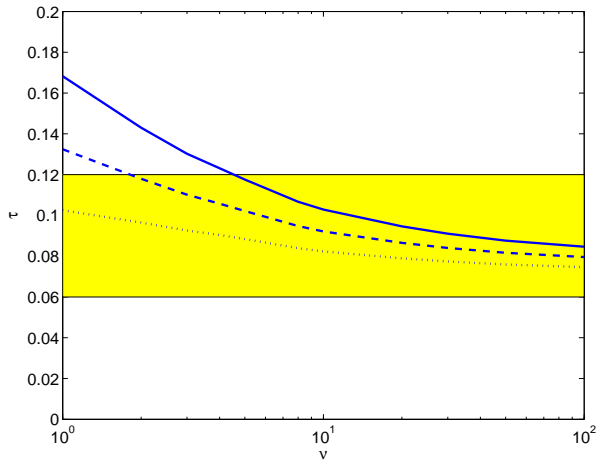,width=8cm}
\caption[Figure5]{\label{f:figure5} The integrated optical depth (up
to $z=60$) due to Thomson scattering, $\tau$, as a function of the
number of $\chi^2$ degrees of freedom considering different evolutions
for the clumping factor parametrized by $\beta=2$ (solid line),
$\beta=1$ (dashed line) and $\beta=0$ (dotted line) and a reionization
model with $\epsilon_{\rm II}=0$. The horizontal stripe represents the
range of $\tau$ allowed by the WMAP three-year data (at $68 \%$
confidence level).}
\end{figure}

\section{Conclusions}

We have investigated the possible impact of non-Gaussian density
perturbations on the reionization history of the Universe in the
context of cosmological models with non-Gaussian density fluctuations
with a renormalized $\chi^2$ probability distribution function
parametrized by the number, $\nu$, of degrees of freedom. We assumed a
constant spectral index so that the amplitude of the density
perturbations on the small scales relevant to reionization could be
calibrated using the three-year WMAP results. We have confirmed
the results of Haiman \& Bryan (2006) suggesting a significant
suppression of star formation in mini-halos. 

We have also shown that the reionization history of the Universe is
able to constrain the non-Gaussian nature of the density
perturbations. However, we have seen that reionization alone is not
yet able to rule out any value of $\nu$ (although very small values of
$\nu$ appear to be disfavoured). Given that large deviations from a
Gaussian probability distribution function on large cosmological
scales are already excluded by the WMAP data, we conclude that unless
the probability distribution function has a strong scale dependence
the cosmic reionization constraints are not competitive with those
coming from the cosmic microwave background anisotropies. Still, it is
important to bear in mind that the scales probed by reionization are
much smaller than those probed by the WMAP (or even Planck), and
consequently the consistency is reassuring, in particular taking into
account recent claims (Mathis, Diego \& Silk 2004) of hints for
non-Gaussianity on scales much larger than those relevant for
reionization.

We should also emphasize that our assumption of a constant spectral
index excludes an important class of models where the density field is
the sum of Gaussian and non-Gaussian contributions with different
power spectra. On large cosmological scales the non-Gaussian component
is severely constrained and only very small deviations from
Gaussianity are allowed. However, if the power spectrum of the
Gaussian contribution is steeper than that of the non-Gaussian one, as
happens in hybrid models with both inflationary and defect
perturbations, the non-Gaussian part may be the dominant component on
small scales while being completely negligible on large cosmological
scales.

Avelino \& Liddle (2004) have recently shown that the
reionization history of the Universe leads to stringent constraints on
the energy scale of defects (see also Olum \& Vilenkin 2006). However,
since in that case the redshift at which the Universe becomes fully
ionized no longer fixes the efficiencies of type I halos, the analysis
is more dependent on the astrophysics of reionization. We shall return
to this issue in a forthcoming publication.

\section*{ACKNOWLEDGMENTS}

P.P.A.\ was supported by Funda{\c c}\~ao para a Ci\^encia e a Tecnologia 
(Portugal) in the framework of the POCTI program, and A.R.L.\ by 
PPARC (UK). A.R.L.\ thanks Jerry Ostriker for a useful discussion.

%\bsp

\end{document}